\documentclass[conference]{IEEEtran}
\IEEEoverridecommandlockouts
\usepackage{cite}
\usepackage{amsmath,amssymb,amsfonts}
\usepackage{algorithmic}
\usepackage{graphicx}
\usepackage{textcomp}
\usepackage{xcolor}
\def\BibTeX{{\rm B\kern-.05em{\sc i\kern-.025em b}\kern-.08em
    T\kern-.1667em\lower.7ex\hbox{E}\kern-.125emX}}
\begin{document}

\title{A Data-Driven Approach for Electric Vehicle Powertrain Modeling\\
}

\author{\IEEEauthorblockN{Eymen Ipek}
\IEEEauthorblockA{\textit{Electrics/Electronics \& Software} \\
\textit{Virtual Vehicle Research Gmbh}\\
Graz, Austria \\
eymen.ipek@v2c2.at}
\and
\IEEEauthorblockN{Assoc.Prof.  Mario Hirz}
\IEEEauthorblockA{\textit{Institute of Automotive Engineering	} \\
\textit{Graz University of Technology}\\
Graz, Austria \\
ORCID: 0000-0002-4502-4255}}

\maketitle

\begin{abstract}
Electrification in the automotive industry and increasing powertrain complexity demand accelerated, cost-effective development cycles. While data-driven models are recently investigated at component level, a gap exists in systematically integrating them into cohesive, system-level simulations for virtual validation. This paper addresses this gap by presenting a modular framework for developing powertrain simulations. By defining standardized interfaces for key components-the battery, inverter, and electric motor-our methodology enables independently developed models, whether data-driven, physics-based, or empirical, to be easily integrated. This approach facilitates scalable system-level modeling, aims to shorten development timelines and to meet the agile demands of the modern automotive industry.
\end{abstract}


\section{Introduction}

Global sales of electric vehicles (EVs) increased by 25\%, reaching 17.7 million units, with battery electric vehicle demand rising by 14\%, plug-in hybrid electric vehicle (PHEV) demand surging by 50\% in 2024, and the global EV fleet projected to grow twelve-fold by 2035 under stated policies \cite{IEA2024, Volta2024}.  

An electric powertrain comprises the system of parts in an EV that identifies a vehicle as an EV, creates and delivers power to make the vehicle move. It comprises a mix of electrical and mechanical components. Powertrain development for electric vehicles begins with defining vehicle-level attributes, such as range, acceleration, efficiency, and thermal management, which are then translated into specific powertrain system requirements. These system-level requirements are further decomposed into individual component specifications \cite{Hofs2018} for motor, inverter and battery. Virtual validation is essential in this process, enabling engineers to simulate how these components interact to achieve the overall system objectives.

Modelling and simulation are the backbone of virtual validation, enabling iterative refinement to verify that the motor delivers the required torque and speed, the inverter operates efficiently under load, and the battery provides sufficient energy, power and thermal stability. This approach not only ensures alignment between vehicle attributes and component performance but also reduces the need for costly physical prototypes and accelerates the development timeline.

Multi-physics-based modelling is a comprehensive approach that spans multiple scales, from the material level to the component level, enabling a deeper understanding of electric vehicle powertrains. At the material level, it captures the physical nature of components such as the electrochemical processes in battery cells \cite{DFN93,Zhang00, Kang20}, the magnetic properties of motor materials \cite{Ismagilov24}, the mechanical \cite{Karthika25} or thermal behavior of motor \cite{Garud25} or characteristics of inverter substrates \cite{Bai23}. The main problems with multi-physics modelling are high computational complexity and resource requirements  \cite{Bai23} resulting from the need to simulate interactions across multiple physical domains and scales. Consequently, although multi-physics-based modelling is useful at the component level, it is not practical to use at the system level due to the computational burden \cite{Bai23}.

Behavioral models are simplified representations of system or component behavior that focus on capturing the essential dynamics and performance characteristics \cite{Bai23,Pillai24, Zhou21, Torregrosa2024, Ferreira24, Carne22, Yan21}, often at a higher level of abstraction, while disregarding detailed physical phenomena. These models are typically reduced-order approximations to describe how components respond under various operating conditions. They do not simulate the underlying physical processes but instead model the observed behavior over a range of conditions, often using mathematical functions \cite{Liu24}. These models are much less computationally demanding. They allow for faster simulations, which are especially valuable when system-level performance needs to be assessed. However, the trade-off is that behavioral models may sacrifice fidelity and accuracy compared to detailed multi-physics simulations \cite{Bai23}. 

Look-up tables or empirical models use pre-measured data or observations to create tables or functions mapping inputs to outputs \cite{Pillai24}. They are computationally efficient and simple \cite{Pillai24}, but limited by range and quality of the data used to generate them \cite{Zhou21}.

Data-driven models rely on machine learning or statistical techniques to identify patterns and relationships within large datasets \cite{Aykol21, Alamin22, Li22, Glucina23, Liu24, Krishna19, Wang24}. They offer flexibility and high adaptability \cite{Ibrahim23, Ibrahim24}, but require significant amount and quality of data for training and validation \cite{Ibrahim23, Wang24}, and their computational demand depends on the complexity of the model \cite{Alamin22}.

The automotive industry is facing a paradigm shift similar to that of "NewSpace" \cite{NewSpace}. Just as NewSpace redefined the aerospace sector through private innovation, agility, and commercial ambition, the "NewAutomotive" era is transforming transportation with electric drivetrains, software-defined vehicles, and tech-driven mobility ecosystems. Like NewSpace, "NewAutomotive" faces the challenge of needing flexible, rapid, and cost-efficient development processes. To meet these demands, increasing modularity and customization, maximizing the reuse of subsystems, iterating on existing solutions, and relying on existing certification pathways are essential.

At the core of this transformation, there is a shift toward data-driven development, which is becoming a new lingua franca of modern engineering. Parametric design and simulation can now be dramatically accelerated through data-driven modeling techniques. This enables highly parallelized virtual validation workflows, made possible by the advanced cloud computing capabilities of modern systems. As a result, multiple environmental conditions and use cases can be virtually assessed across a wide range of vehicle architectures, powertrain systems, and component configurations. Therefore, accelerated simulation enables shortening development time of the control system software. 

\begin{figure}[h!]
\centerline{\includegraphics[width=\linewidth]{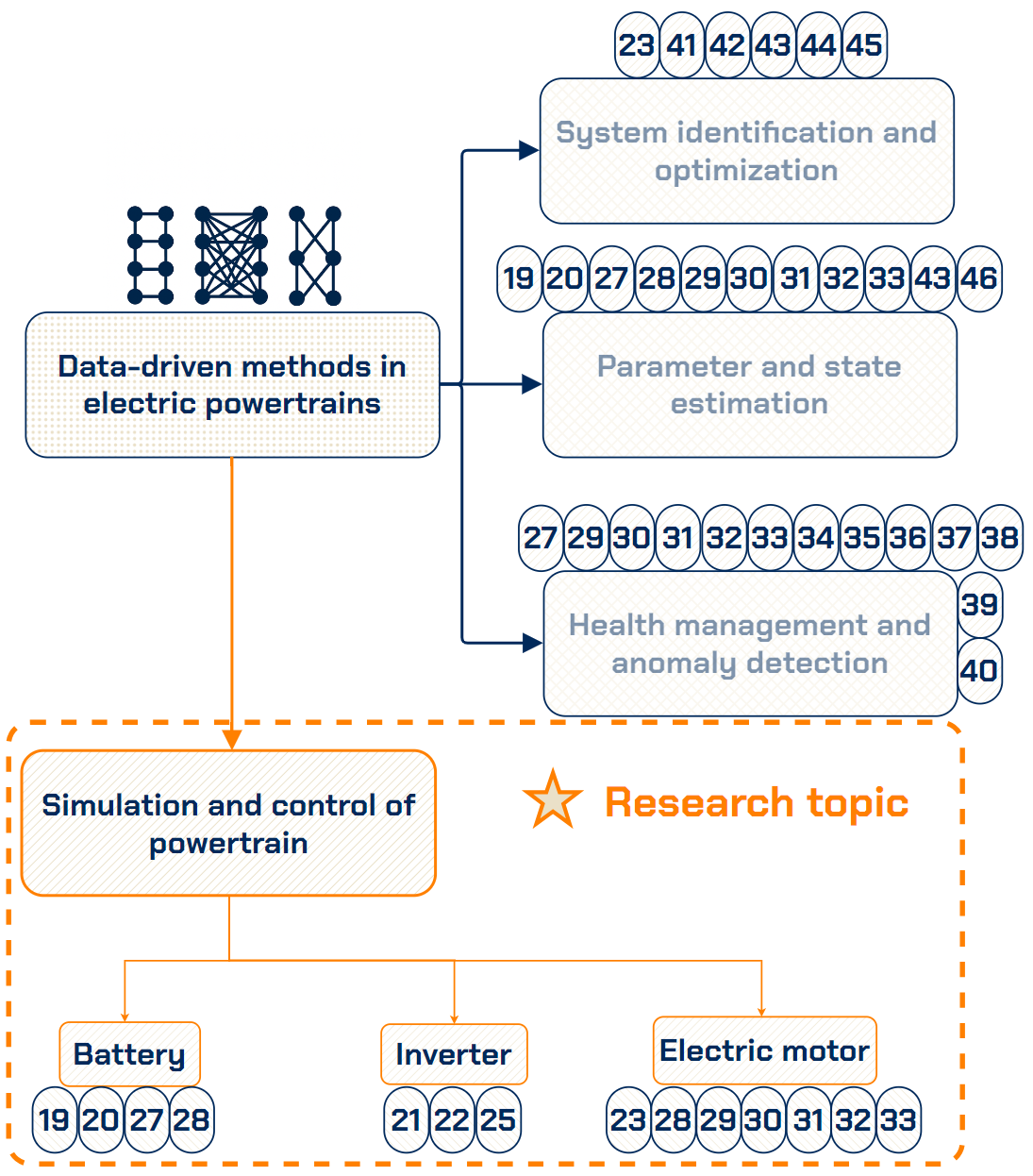}}
\caption{Literature review and research gap. Although there are data-driven studies focusing on data-driven modeling and simulation for individual components, this research targets an integrated powertrain simulation composed by these individual components.}
\label{fig1}
\end{figure}

As illustrated in Fig. \ref{fig1}, although there are data-driven studies focusing on state estimation \cite{Mayemba24,Alamin22, Hagenbucher25, Bouziane24, Kirchgaessner21, Nawae20, Mukherjee20, Mukherjee22}; prognostics and health management \cite{Yao21, Cai17, Senanayaka18, Park22, Liu21, Peng22, Ulatowski16}; consumption optimization \cite{Lei24} and energy management \cite{Marchand23}; to predict temperature \cite{Hagenbucher25, Bouziane24, Kirchgaessner21, Okada20}; to estimate torque \cite{Bouziane24,Nawae20, Mukherjee20, Mukherjee22} and speed \cite{Mukherjee20, Mukherjee22}; for optimizing \cite{Sakamoto23} or predicting motor performance \cite{Wang24, Manohar23} ; or even for creating digital twin instances based on real-world fleet data \cite{Hagenbucher25}; all these approaches are out of scope of this study. Unlike existing studies that focus on isolated subsystems, this work presents a unified framework for integrating individual data-driven electric powertrain components into a cohesive system. This approach uniquely combines classical physics-based models and data-driven surrogates, managed by rule-based controllers. By establishing this hybrid methodology, the work bridges a critical gap in the literature, enabling the synchronization of deterministic logic and data-driven behaviors.

\section{Methodology}
Design and optimization of EV powertrains increasingly rely on advanced modeling techniques. As summarized in Fig. \ref{fig1}, the existing body of research shows a strong focus on applying these methods to isolated challenges.

However, a critical limitation of this component-centric approach is that it overlooks the complex interactions between different parts of the powertrain. The performance of the motor, for instance, is directly dependent on the quality of the power delivered by the inverter, which in turn is constrained by the battery's voltage and current limits. To capture these system-level dynamics, a holistic and integrated simulation is essential. This research, therefore, focuses on developing an integrated powertrain simulation framework that unites individual models of the battery, inverter, and electric motor, allowing for a comprehensive analysis of the entire system's behavior.

\begin{figure}[h!]
\centerline{\includegraphics[scale=0.13]{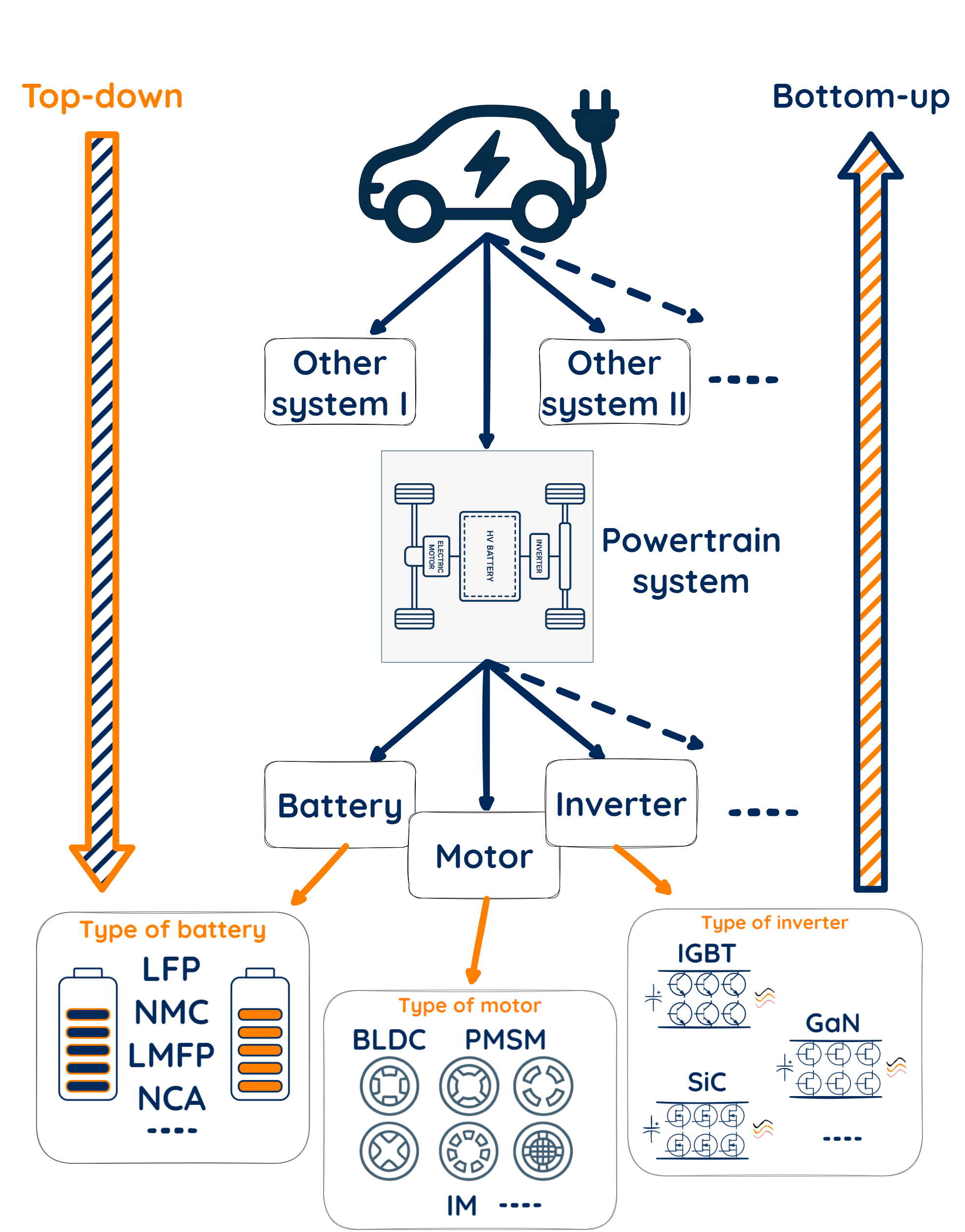}}
\caption{Structure of model development. The proposed methodology suggests starting with higher-level systems by defining their required inputs and outputs. Then, a top-down approach is used to decompose the system to the smallest modelable element. Finally, a bottom-up approach is followed to integrate them.}
\label{fig2}
\end{figure}

To manage the complexity of building this integrated model, we propose a structured modeling framework based on the well-established V-model, as shown in Fig. \ref{fig2}. This process ensures both logical decomposition and robust integration.

The process starts with a top-down decomposition for requirements specification. We begin at the highest level and progressively break the system down into its core subsystems. The powertrain system is then decomposed into its fundamental components: the battery, inverter, and motor. This decomposition continues until we isolate the smallest element to be modeled, defining not just the components but also their specific types. This decomposition is vital for precisely defining the required inputs, outputs, and interfaces for each model.

Following the development of individual component models, we employ a bottom-up integration approach for verification and validation. The models are progressively combined, and at each stage, the integrated system is tested to ensure it functions as expected. This methodical assembly guarantees that the final powertrain model is a faithful representation of the real-world system.

The proposed approach suggests creating, training, and validating component models in Python\cite{Python23} using well-established artificial intelligence libraries. Then, validated models can be used in Simulink\cite{Simulink24} for integration at system level. Therefore, the framework is backward-compatible and can be used with rule-based controllers. If necessary, the data driven plants with their controllers can be exported to Functional
Mock-up Unit. Furthermore, existing physics-based or behavioral models in X-in-the-loop setups can be replaced by data-driven models to accelerate the simulation.

\begin{figure}[h!]
\centerline{\includegraphics[scale=0.3]{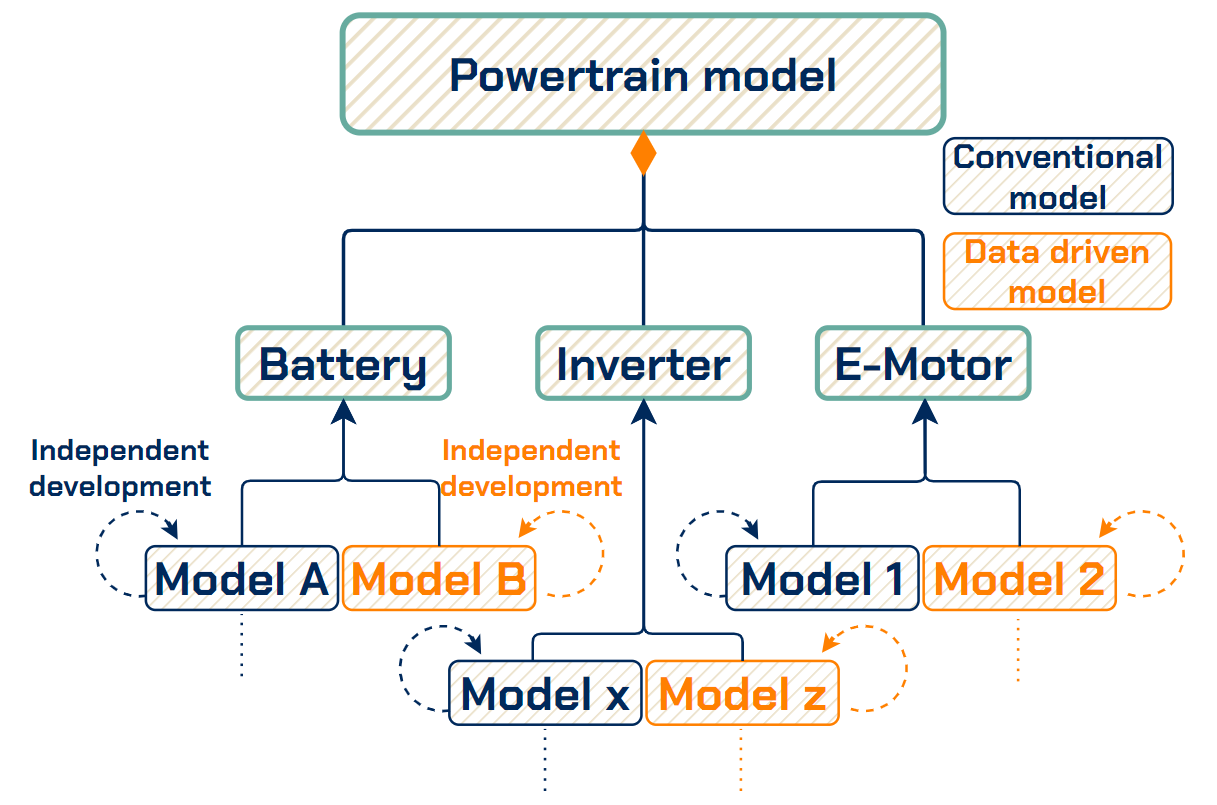}}
\caption{Architecture of model development. The proposed methodology can independently integrate conventional physics-based and data-driven component models into a cohesive powertrain simulation.}
\label{fig3}
\end{figure}

The cornerstone of our methodology is a flexible and modular architecture that supports hybrid modeling, as depicted in Fig. \ref{fig3}. This framework is designed to integrate data-driven and conventional models-such as physics-based, behavioral, or empirical-interchangeably.

Our architecture allows a developer, for example, to pair a highly accurate physics-based model for a novel battery chemistry (Model A) with a computationally efficient data-driven model for a standard, well-characterized electric motor (Model 2). This "mix-and-match" capability enables the creation of a simulation that is uniquely tailored to specific design goals, balancing accuracy, computational speed, and development effort.

With this approach, it is possible to use augmented data from component-level, high-fidelity simulations or empirical data gathered from component-level testing on dynamometers and cell cyclers to train surrogate models that accurately predict physical behavior. These data-driven models replace complex first-principles equations. This approach significantly accelerates system-level simulation time while maintaining high fidelity to the real-world components. The resulting models are computationally efficient, making them exceptionally suitable for extensive design space exploration, Hardware-in-the-Loop (HIL) validation, energy consumption studies over standardized drive cycles, and the development of advanced control strategies.

\begin{figure*}[h!]
\centerline{\includegraphics[scale=0.28]{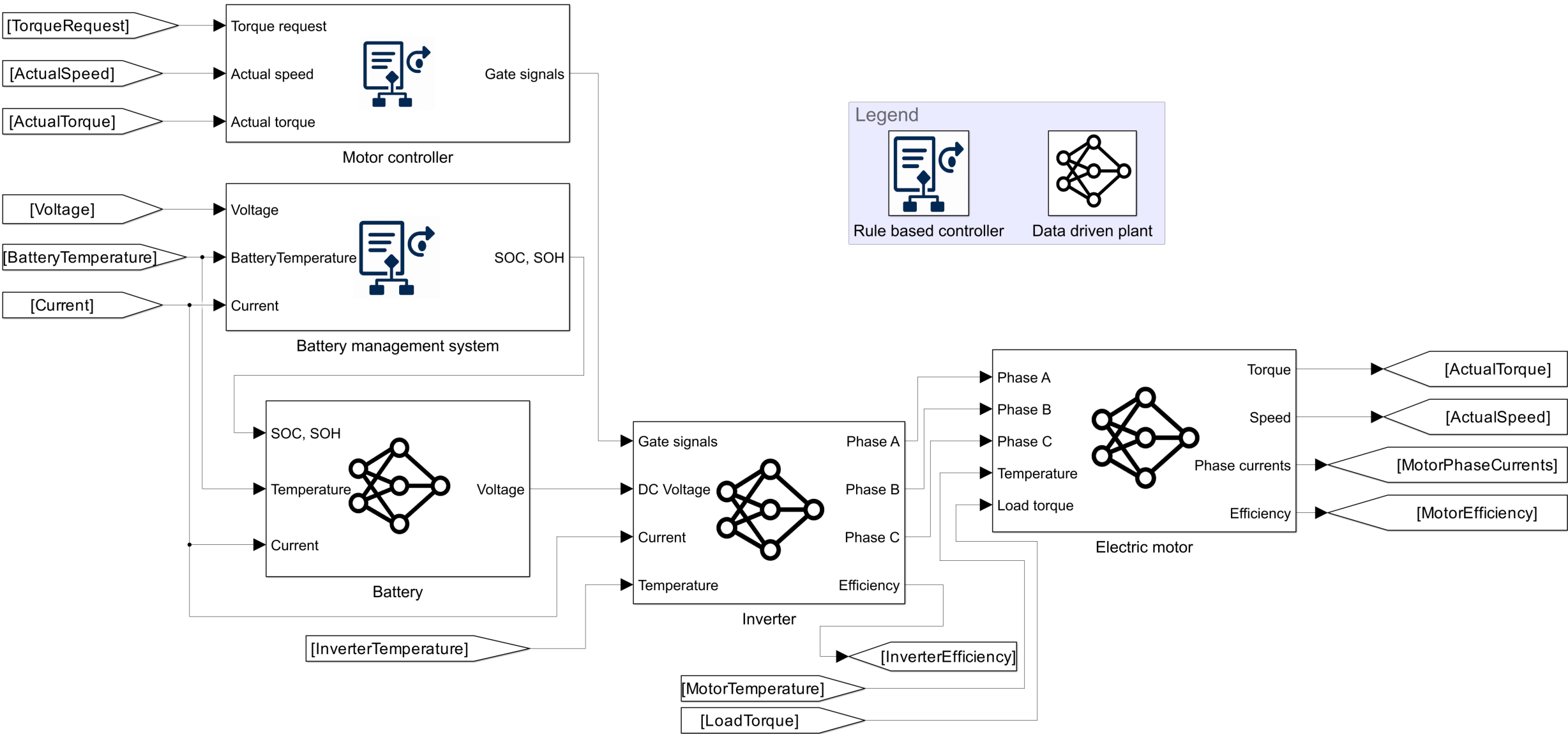}}
\caption{Overview of the proposed framework for an EV drive system. This model integrates traditional rule-based logic for control components with data-driven plant models for the core powertrain. These plant models use artificial intelligence to accurately capture complex nonlinear behaviors, estimating the desired outputs.}
\label{fig4}
\end{figure*}

A practical implementation of this integrated framework is shown in the block diagram in Fig. \ref{fig4}. The model shows the combination of rule-based controllers with pre-trained, data-driven plant models. The models can be developed independently under various conditions. Furthermore, each plant model acts according to the inputs from adjacent plants and the controllers. As the approach is based on data-driven plants, the signal flow consists of information rather than physical signals. To achieve a comprehensive system representation, a dedicated thermal model must be developed in parallel. This thermal model is not merely a simple extension of each electrical component but a distinct system that must be separately considered and trained to reflect the specific thermal architecture of the vehicle, including its cooling circuits and heat exchange pathways.

The development process for each plant model is systematic and data-centric. For the battery, a dynamic model is trained to predict terminal voltage by characterizing its open-circuit voltage and internal resistance as functions of current draw, state of charge, and temperature. The inverter model is designed to output 3-phase AC voltages based on the DC-link voltage it receives from the battery and a simplified representation of the controller's commands, such as modulation index and frequency. Finally, the electric motor model is characterized across its full torque-speed map. It takes the inverter's synthesized AC voltages and an external load torque as inputs to predict the resulting shaft torque, rotor speed, and, critically, the AC currents drawn from the inverter.

For system integration within a simulation environment, these open-loop models are connected sequentially. A rule-based controller closes the primary control loop by processing the driver's torque request and motor feedback to generate commands for the inverter. A crucial energy-balance feedback path is the calculation of the battery's load. The instantaneous AC power consumed by the motor is calculated from its voltages and currents. This power is then adjusted using a data-driven efficiency map of the inverter to determine the required DC power. The resulting DC current is then fed as the load input to the battery model, ensuring a physically coherent power flow and enabling a robust, system-level simulation.

\section{Results and discussion}

The development of this hybrid modeling framework demonstrates a significant step toward creating more holistic and adaptable powertrain simulations. The primary strength of the proposed architecture lies in its modularity, which allows for a strategic balance of model fidelity trade-offs. Engineers can select high-fidelity, physics-based models for components requiring detailed analysis, such as a new battery chemistry under thermal stress, while employing computationally efficient, data-driven models for well-understood components like the inverter. However, this flexibility introduces challenges, most notably the data dependency and generalization issues inherent in machine learning models. The accuracy of a data-driven component is fundamentally limited by the scope and quality of its training data. A model trained exclusively on standard drive cycles, for example, may fail to accurately predict performance during aggressive driving or in extreme temperatures, highlighting the critical need for comprehensive and diverse datasets to ensure robust generalization.

Looking forward, future work will focus on validating and extending the practical applicability of this framework. A next step is to apply this methodology to a concrete example, such as simulating the powertrain for a specific vehicle platform to benchmark its performance against real-world test data. This will provide a clear validation of the integrated model's predictive accuracy. Beyond validation, there is significant potential in pursuing a multi-objective optimization of the different machine learning methods used within the framework. This would involve systematically tuning algorithms to find the optimal compromise between predictive accuracy, computational cost, and memory usage, which is essential for eventual deployment in real-time control systems. Ultimately, the long-term goal is a thorough evaluation of real-world usage and practical applicability, likely through HIL simulations, to rigorously assess the framework's performance and reliability under the unpredictable conditions of real-world operation.

\section{Conclusion}

This research has outlined a comprehensive and flexible framework for the integrated simulation of electric vehicle powertrains. By combining a structured top-down decomposition and bottom-up integration methodology with a modular architecture, our approach unites conventional physics-based models with modern data-driven techniques. The key contribution of this work is the ability to strategically manage the trade-offs between model fidelity and computational cost, allowing developers to tailor simulations to specific design and analysis needs. While acknowledging challenges such as data dependency and the need for real-world validation, this hybrid modeling paradigm establishes a robust foundation for future research. It paves the way for more accurate system-level analysis, multi-objective optimization, and the accelerated development of next-generation powertrain control strategies, ultimately contributing to the advancement of electric vehicle technology.

\section*{Acknowledgment}
This work was funded by the program “Industrienahe Dissertation 2024” of the Austrian Federal Ministry for Innovation, Mobility and Infrastructure (BMIMI). 
The publication was written at Virtual Vehicle Research GmbH in Graz and partially funded within the COMET K2 Competence Centers for Excellent Technologies from the Austrian Federal Ministry for Innovation, Mobility and Infrastructure (BMIMI), Austrian Federal Ministry for Economy, Energy and Tourism (BMWET), the Province of Styria (Dept. 12) and the Styrian Business Promotion Agency (SFG). The Austrian Research Promotion Agency (FFG) has been authorised for the programme management. Authors furthermore express thanks to the supporting industrial partner AVL List GmbH.

  \centering
  \includegraphics[scale=0.11]{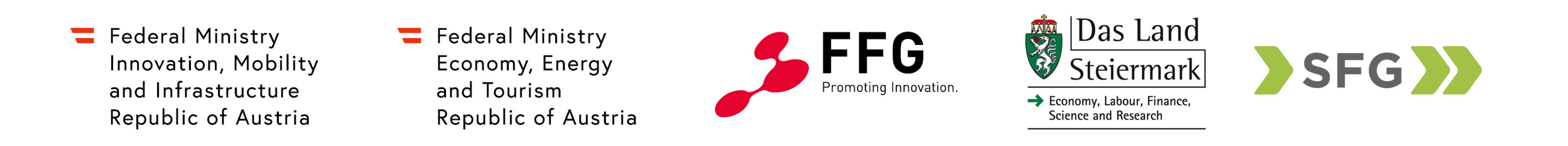} 



\vspace{12pt}
\color{red}
\end{document}